\documentclass[aps,pra,preprint,groupedaddress]{revtex4-1}
\usepackage{}
\usepackage{amsfonts}

\usepackage{graphicx}
\usepackage{dcolumn}
\usepackage{bm}
\usepackage{amssymb}
\usepackage{amsthm}
\usepackage{amsmath}
\usepackage{amsfonts}
\usepackage{multirow}
\usepackage{CJK}
\renewcommand{\vec}[1]{\mbox{\boldmath$#1$}}

\begin{document}

\begin{CJK*}{GB}{gbsn}
\title{Quantum Measurement and Extended Feynman Path Integral}
\author{Wei Wen (ÎÄΰ)$^{\ast 1}$, Yan-Kui Bai (°×Ñå¿ý)$^2$}
\affiliation{$^{1}$Institute of Semiconductors, Chinese Academy of Sciences,
Beijing, China}
\email{chuxiangzi@semi.ac.cn}
\affiliation{$^{2}$College of Physical Science and Information Engineering, Hebei Normal University, Shijiazhuang, Hebei 050016, China}

\begin{abstract}
Quantum measurement problem still is unconsensus since it has existed many years and inspired a large of literature in physics and philosophy.
We show it can be subsumed into the quantum theory if we extend the Feynman path integral by considering
the relativistic effect of Feynman paths. According to this extended theory, we deduce not only
the Klein-Gordon equation, but also the wave-function-collapse equation. It is showing that the stochastic and instantaneous collapse of the quantum measurement is due to the ``potential noise'' of the apparatus or environment and ``inner correlation'' of wave function respectively. Therefore, the definite-status of the macroscopic matter is due to itself and this does not disobey the quantum mechanics. This work will give a new recognition for the measurement problem.
\end{abstract}

\maketitle
\end{CJK*}
\section{Introduction}
Quantum measurement is a old and vexed topic in quantum world. It is brought into quantum theory in order to solve the conflict between the definite status in macroscopic world and superposition-state in microscopic world. According to the standard interpretation of quantum measurement, the definite status is due to the superposition state collapsed under a measurement. It seems that the micro-to-macro conflict is solved under the standard quantum theory interpretation, but it brings out more arguments than that it solves because this interpretation does not give the further instruction about why and how this collapse happens \cite{qmp1,qmp2,qmp3,qmp4,qmp5,qmp6,qmp7,qmp8,qmp9,qmp10,qmp11,qmp12,qmp13,qmp14,qmp15,qmp16,qmp17,qmp18}. These arguments focus on the definite outcomes problem \cite{dfo1,dfo2,dfo3,dfo4}, prefer-basis problem \cite{pfsp1,pfsp2,pfsp3,pfsp4,pfsp5} and the probability and instantaneity character of the wave function collapse \cite{qnl1,qnl2,qnl3,qnl4,qnl5}. Especially, the process of quantum measurement, whether the character instantaneity and probability could be resumed into the present theory is unknown yet, because the former seems to break the special relativity theory, and the latter appears to incompatible with the reversible time evolution of the Schr\"{o}dinger equation. Therefore quantum measurement becomes one of the most puzzled concepts in QM .

In this paper, we propose a new theory to review the physical mechanism of the quantum collapse. It is called extended Feynman path integral (EFPI). This new theory not only is compatible with the standard quantum theory, but also could deduce some new conclusion, such as the the Klein-Gordon equation but also show the inner nonlocal correlation of wave function. Moreover, this new theory could be used to clearly and perfectly interpret the unsolved problems of quantum measurement proposed above. The EFPI is the a relativity modification of Feynman path integral (FPI).  In fact, FPI is a non-relativistic formulation because the Lagrangian action $S=\int_0^{t}L(\tau)\tau$ is a non-relativistic \cite{Fyman}. Therefore, it is necessary to consider the a relativity extension of FPI. It is worth reminding that this extension is different from the quantum field theory because we modify the formula of FPI rather than quantize  the wave function. According this extension, we get a collapse equation, showing that the stochastic and instantaneous collapse of the quantum measurement is due to the ``potential noise'' of the apparatus or environment and ``inner correlation'' of wave function. Moreover, to test and verify these conclusions, we design a suggestive experiment, which is based on the magnetic Aharonov-Bohm effect \cite{ABE1,ABE2,ABE3,ABE4}.

The novelty of this work is it subsumes quantum measurement into quantum theory and predicts the quantum measurement can be control. It not only perfectly demonstrates the probability and instantaneity character of the wave function collapse, and solves the preferred-basis problem in quantum measurement, but also firstly shows the origin of the quantum nonlocality in wave function.  Theoretically, it is a break-through for the nowadays quantum theory and can be seemed as an alternate way to solve the other problems and assumptions in quantum theory, such as the spatio-temporal view of the teleportation in quantum information and the non-preference assumption in statistical mechanics.

This article will be organized into five parts. First of all we discuss why and how to extend FPI. In this section, a new formalization of path integral is obtained. It is shown its low-energy approximation is FPI. In next section, we point out that the Klein-Gordon equation is just a deduction of EFPI. However, the EFPI is more fundamental than the time-evolution equation. As can be seen the Klein-Gordon equaiton cannot describe the single particle's evolution but the EFPI can. We also explore density-flux equation and the nonlocal correlation of wave function. In section 4, we come to the other core of the paper by interpreting how the wave function collapse happens though the EFPI. Schr\"odinger cat paradox is detailed and the numerical simulation of two-state collapse is given. Subsequently, we detail a thought experiment to test and verify the EFPI theory. We predict that the interference effect will transform into the diffraction effect in the magnetic Aharonov-Bohm experiment if some special fluctuated magnetic field is added in. Finally, we will conclude our analysis with a short discussion of the relationship between EFPI and other interpretations.

\section{The Extended Feynman Path Integral}
In the history of the quantum theory, there are three well known expressions, namely the differential equation of Schr\"odinger, the matrix algebra of Heisenberg and the FPI. These three expressions always seemed equivalent, but they have their own focuses: The Schr\"odinger and Heisenberg expressions focus on the evolution of states and operations respectively, whereas, the path integral formulation of Feynman on the ``relation'' between Lagrangian mechanics and quantum mechanics. FPI has proved crucial to the subsequent development of theoretical physics, especially of the quantum field and quantum statistics. In this section, we show it is also the candidate in solving the quantum measurement problem.

\subsection{The Relation Between FPI and Wave Function Collapse}
It is known that FPI is extended theory for the quantum propagator, describing
the ``correlation'' of point to point as states are evolving and the another ``correlation'' in quantum mechanics appears in the process of quantum measurement. When a quantum measurement is made on a wave function diffusing in all of space, such as the measurement of the position of an
electron in the experiment of double-slit interference, it will be found that
the whole wave function will instantaneously change and collapse into the measured
position with the exact location described as a range of probabilities. It is thought that
there is some inner ``correlation'' in the wave functions transferring the
action of the measurement from the local part to the whole. Comparing these two
``correlations'', once the process of wave-function collapse can be subsumed
into quantum dynamics by some theory, it might be the Feynman path integral.

Furthermore, the action in Feynman path integral
formulation is a non-relativistic form \cite{ftnt3}. The question arisen is: what would happen when the non-relativistic action is extended to the relativistic action? According to the thought of Feynman, all possible paths should be included to compute the propagator in non-relativistic quantum mechanics. Therefore, the superluminal trajectories might be included in possible paths to calculate the quantum amplitude. And then, we want to calculate the relativistic effect of the trajectories of FPI.

\subsection{The Concrete Form of The EFPI}
The propagator in non-relativistic quantum mechanics is defined as
\begin{equation}
K(\vec{r},\vec{r}_{0};t,t_{0})=\langle \vec{r}|\hat{U}(t,t_{0})|\vec{r_{0}}%
\rangle ,  \label{Propagator}
\end{equation}%
where $\hat{U}(t,t_{0})$ is the unitary time-evolution operation for the
system taking states at time $t_0$ to the states at time $t$: $
\hat{U}(t,t_{0})|\Phi (t_{0})\rangle =|\Phi (t)\rangle $.

In 1984 Richard Feynman found a space-time approach that is proportional to the
propagator in quantum mechanics. It is known as the FPI. In this theory, the propagator in quantum theory can be expressed as
\begin{equation}
K(\vec{r},\vec{r}_0;t,t_0)=C\sum_{all\ \ paths}\exp(\mathrm{i}S/\hbar),
\label{Fymeq}
\end{equation}
According to the second postulate of Feynman \cite{Fyman}, the coefficient $C$  in Eq.~(\ref{Fymeq}) is considered to be a constant independent of the paths. However, is it necessary for $C$ to be a constant? or is this just an approximation of a higher-level theory in some conditions such as when relating Newtonian mechanics to the relativity? With a bold try, we rewrite the Eq.~(\ref{Fymeq}) into a general form as following (To avoid confusion, $F(\vec{r},\vec{r}_0;t,t_0)$ is used here).
\begin{equation}
F(\vec{r},\vec{r}_0;t,t_0)=R(\vec{r},t_0)\sum_{all\ \ paths}W(\wp)*\exp(\mathrm{i}S/\hbar),
\label{BasicEq}
\end{equation}
where, the path functional $W(\wp)$ is the path weight factor
and the coefficient function $R(\vec{r},t_0)$, independent of paths, is the whole weight factor (be notified that the form $R(\vec{r},t_0)$ depending on the structure of the space manifold, viz., the dimensionality and continuity, can be a real function or differential operator). Basing on our analysis, The formulation of $F(\vec{r},\vec{r}_0;t,t_0)$ should be satisfied three limitations:

\begin{enumerate}
  \item The mathematical requirements. A proper formula about $F(\vec{r},\vec{r}_0;t,t_0)$ should be satisfied the combination rule $F(\vec{r},\vec{r}_0;t,t_0)=\int{F(\vec{r},\vec{r_1};t,t_1)F(\vec{r_1},\vec{r}_0;t_1,t_0)d^3\vec{r_1}}$. and the summation in Eq.~(\ref{BasicEq}) (for the continuous spectrum, the summation transforms into integral formula) should  convergence.
  \item The physical requirements.  A proper formula about $F(\vec{r},\vec{r}_0;t,t_0)$ should be consisted of four-dimension scalars, vectors or tensors.
  \item simplification and compatibility. A proper formula about $F(\vec{r},\vec{r}_0;t,t_0)$ should be as simple as possible and can transform into the formula $K(\vec{r},\vec{r}_0;t,t_0)$ in non-relativistic theory.
\end{enumerate}

It should be reminded that the first limitation is the guarantee for us to get the time-evolution equation. To show this conclusion, we should use the character of the propagator $F(x,x_0;t_0+\varepsilon,t_0)$, where $\varepsilon\rightarrow0$ and calculate the $\phi(x,t_0+2\varepsilon)$ using two approaches. On one side,
\begin{eqnarray}
&&\phi(x,t_0+2\varepsilon)=\int  F(x,x_0;t_0+2\varepsilon,t_0)\phi(x_0,t_0)d x_0 \nonumber \\
\Rightarrow &&\phi(x,t_0)+\frac{\partial \phi(x,t_0)}{\partial t_0}(2\varepsilon)=\phi(x,t_0)+\hat{G}(2\varepsilon)\phi(x,t_0);
\end{eqnarray}
On other side,
\begin{eqnarray}
&&\phi(x,t_0+2\varepsilon)=\int  F(x,x_0;t_0+\varepsilon,t_0)\phi(x_0,t_0+\varepsilon)d x_0 \nonumber \\
&& \ \ \ \ \ \ \ \ \ \ \ \ =\int F(x,x_0;t_0+\varepsilon,t_0)\int F(x,x_1;t_0+\varepsilon,t_0)\phi(x_1,t_0)dx_1dx_0 \nonumber\\
\Rightarrow &&\phi(x,t_0)+\frac{\partial \phi(x,t_0)}{\partial t_0}(2\varepsilon)=\phi(x,t_0)+2\hat{G}(\varepsilon)\phi(x,t_0).
\end{eqnarray}
To obtain these two equation, we should note $\lim_{\varepsilon\rightarrow 0} \hat{G}(\varepsilon)=0$ and $\lim_{\varepsilon\rightarrow 0}{ \hat{G}(\varepsilon)}/{\varepsilon}=cons.$. Therefore, according these two conclusion, if $F(\vec{r},\vec{r}_0;t,t_0)$  satisfies the combination rule, then,
\begin{eqnarray}
&&\hat{G}(2\varepsilon)=2\hat{G}(\varepsilon) \Leftrightarrow \hat{G}(\varepsilon)=\hat{g}\varepsilon; \label{eq1}  \\
\Rightarrow&&\frac{\partial \phi(x,t_0)}{\partial t_0}=\hat{g}\phi(x,t_0). \label{eq2}
\end{eqnarray}
Comparatively, if the $F(\vec{r},\vec{r}_0;t,t_0)$ does not satisfy the first limitation, then $\hat{G}(\varepsilon)$ cannot be rewritten as the separation of variable $\varepsilon$, namely $ \hat{G}(\varepsilon)\neq\hat{g}\varepsilon$ and therefore, we could get no time-evolution equation.

These limitations demarcate the formula of $F(\vec{r},\vec{r}_0;t,t_0)$. For example, according to these three limitations, the $W(\wp)$ cannot be a constant, or else, on the assumption that $W(\wp)=W_0$, we will get the infinitesimal time-evolution as
\begin{equation}
  F(x,x_0;t_0+\varepsilon,t_0)=R(x,t_0)W_0\exp\left(\mathrm{i} {L\left((x_0+x)/2,\frac{x_0-x}{\varepsilon}, t_0\right)}/{\hbar}\varepsilon\right),
  \label{minitime}
\end{equation}
where $L(x,\dot{x},t)$ is Lagrangian function in relativity. Consequently,
\begin{eqnarray}
&&\phi(x,t_0+\varepsilon) \nonumber\\
&&=R(x,t_0)W_0\int_{-\infty}^{\infty}{\exp\left(\mathrm{i} L({(x_0-x)}/{\varepsilon})\varepsilon/{\hbar}\right)\phi(x_0,t_0)dx_0} \nonumber\\
&&=R(x,t_0)W_0\int_{-\infty}^{+\infty}{\exp\left(-\mathrm{i}\sqrt{1-v^2/c^2}\varepsilon/\tau_0\right)\phi(x+v\varepsilon,t_0)\varepsilon dv} \nonumber \\
&&=R(x,t_0)W_0\phi_0\pi c \varepsilon \left(-Y_1(\varepsilon/\tau_0)+\mathrm{i} J_{1}(\varepsilon/\tau_0)\right),
\label{odst}
\end{eqnarray}
where, $J_{v}(z)$ and  $Y_{v}(z)$ are called the Bessel functions of first kind and the Bessel functions of the second kind or Weber's
function respectively, and $\tau_0=\hbar/(mc^2)$. At the last step of Eq~(\ref{odst}), we set $\phi(x_0,t_0)=\phi_0$. Because the Eq~(\ref{odst}) should be hold when $\varepsilon\rightarrow 0$, therefore, $R(x,t_0)=(2 W_0c)^{-1}$. We can see, $\phi (x_0,t_{0}+\varepsilon )$ can be expanded into the integer-order series of $\varepsilon$
\begin{equation}
\phi (x_0,t_{0}+\varepsilon )=\phi(x_0, t_0)+\varepsilon\frac{\partial\phi(x_0,t_0)}{\partial t}+O(\varepsilon^2).
\end{equation}
However, $Y_{1}(\varepsilon/\tau_0)$ can not be expanded into the integer-order series of $\varepsilon$
\begin{eqnarray}
Y_{1}(\varepsilon/\tau_0)&=&-\frac{2\tau_0}{\pi \varepsilon}+\varepsilon\frac{-1+2\gamma_E-2\ln(2)+2\ln(\varepsilon/\tau_0)}{2\pi \tau_0}+O(\varepsilon^3); \nonumber \\
J_{1}(\varepsilon/\tau_0)&=&\frac{\varepsilon}{2\tau_0}+O(\varepsilon^3),
\end{eqnarray}
where $\gamma_E$ is the is Euler's constant,  with numerical value $\simeq0.5772$.  Because $Y_{1}(\varepsilon/\tau_0)$ contains the term $\ln(\varepsilon/\tau_0)$, which it cannot expanded into the integer-order series of  $\varepsilon$, therefore, the combination rule can not be satisfied. Similarly, the formula like $W(\wp)=1/(\Delta \tau)^{n}$, where $n\in Reals$, is also improper because the $F(x,x_0;t,t_0)$ cannot convergence.

Although these three limitations could eliminate many possible candidates formula of $F(\vec{r},\vec{r}_0;t,t_0)$, it is still more inspirational than rational for us to fix on an appropriate formula. Basing on many calculations, we find $W(\wp)$ could be finally explored as
\begin{equation}
W(\wp)=\frac{\mathbb{P}(\wp)}{\mathcal{P}(\wp)}\left(\frac{\triangle\tau}{2}\right)^{-1/2},
\label{fracEq}
\end{equation}
$\mathcal{P}
(\wp)=\int_{t_0}^{t}{\sqrt{2mT}d\tau}$. $P$, $T$ and $\Delta \tau $ are called the momentum, kinetic energy and proper time in terms of four-dimensional space-time. Their formulations are the followings
\begin{eqnarray}
\mathbb{P}(\wp)&=&\int_{t_0}^{t}{|P|d\tau}=\int_{t_0}^{t}\frac{m\dot{x}}{\sqrt{1-\dot{x}^2}/c^2}d\tau;\nonumber\\
\mathcal{P}(\wp)&=&\int_{t_0}^{t}{\sqrt{2mT}d\tau}=\int_{t_0}^{t}\sqrt{2m^2c^2\left(1/\sqrt{1-\dot{x}^2/c^2}-1\right)}d\tau;\nonumber\\
\Delta \tau&=&\int_{t_0}^{t}\sqrt{1-\dot{x}^2/c^2}d\tau.
\end{eqnarray}
This definition of $W(\wp)$ agrees with all the mathematical and physical demands proposed above. In non-relativistic mechanics $\mathbb{P}= \mathcal{P}$ and $\Delta \tau=t_1-t_0$, then the coefficients are independent of paths and the theory transforms to the Feynman's path integral formulation; In the relativity realm, $\mathbb{P}$ and $\mathcal{P}$ are no longer equivalent and new conclusions will be reached. The special differences between EFPI and FPI is shown in Tab.~1.

\section{The Time-evolution equation and Density-flux equation  based on EFPI}
In this section, several new results about some important concepts based on EFPI are obtained. First of all, we should note
\begin{equation}
\phi(\vec{r},t)=\int_{-\infty}^{\infty}R(\vec{r},t_0)d \vec{r}_1\int_{\vec{r_0}}^{\vec{r_1}}D[\vec{r}_2(t)]
W(\wp)\exp\left(\mathrm{i}\int_{t_0}^{t}L(\vec{r}_2,\dot{\vec{r}_2})/\hbar dt_1\right)\phi(\vec{r}_2,t_0),
\label{tevol}
\end{equation}
and the complete formulation of the Lagrangian in relativity
\begin{equation}
L=-m c^2\sqrt{1-\vec{v}^2/c^2}+\vec{A}(\vec{r}_0,t_0)\cdot\dot{\vec{r}}_0-V(\vec{r}_0,t_0),
\end{equation}
of which $A(\vec{r},t)$ and $V(\vec{r},t)$ are the vector potential and the scalar potential respectively. We also should set the rule: $\sqrt{1-\vec{v}^2/c^2}=-\mathrm{i}\sqrt{\vec{v}^2/c^2-1}$ if $|\vec{v}|\geq c$. The trajectories in the FPI are not the real movement of a particle, namely the $\vec{v}$ is not the real velocity of a particle, and therefore $|\vec{v}|$ could be greater than the velocity of light $c$.

\subsection{Time-evolution equation}
The FPI reproduces the Schr\"{o}dinger equation. The theory presented here gives a
new differential equation.
\begin{equation}
\mathrm{i}\hbar\frac{\partial\phi(\vec{r},t)}{\partial t}=\left\{ \sqrt{
m^2c^4+(-\mathrm{i}\hbar\nabla-\vec{A}(\vec{r},t))^2c^2}+V(\vec{r},t)\right\} \phi(%
\vec{r},t),
\label{3DTEEq}
\end{equation}
Eq.~(\ref{3DTEEq}) will be reduced to the Schr\"{o}dinger equation for low energy cases: $\langle\phi| \hat{\vec{p}}^2c^2|\phi \rangle\ll m^2c^4$. One can obtain Eq.~(\ref{3DTEEq}) by using the corresponding relation between operations and physical quantities, such as $\mathrm{i}\hbar\partial_t\rightarrow E $ and $-\mathrm{i}\hbar \nabla\rightarrow \vec{p} $ into the equation $E=\sqrt{m^2c^4+\vec{p}^2c^2}$. In fact, Oscar Klein, Walter Gordon and Paul Dirac got their relativistic quantum equations with the help of the corresponding relation. Here, we show it is just a conclusion of EFPI.

To demonstrate the Eq.~(\ref{3DTEEq}), we give a detail deduction of the free-particle equation in one-dimension in the followings. The other situations can be seen in detail in SOM (Supporting Online Material).

To get the time-evolution equation from the Eq.~( \ref{tevol}), we should consider the infinitesimal time distance $\varepsilon=t-t_0 \rightarrow 0$ in Eq.~( \ref{tevol}). According the analysis of Feynman, the integral of \ref{tevol} can be taken along a straight line for infinitesimality $\varepsilon$. Therefore, we can get the following equations.
\begin{eqnarray}
&&\phi(x,t_0+\varepsilon) \nonumber\\
&&=R(x,t_0)\int_{-\infty}^{\infty}{W\left(x,\frac{x_0-x}{\varepsilon},
\varepsilon\right)\exp\left(\mathrm{i}L(\frac{x_0-x}{\varepsilon})\varepsilon/\hbar\right)\phi(x_0,t_0)dx_0} \nonumber\\
&&=R(x,t_0)\int_{-\infty}^{+\infty}{\varepsilon W(x,v,\varepsilon)\exp\left(-\mathrm{i}\sqrt{1-v^2/c^2}\varepsilon/\tau_0\right)\phi(x+v\varepsilon,t_0)dv} \nonumber\\
&&=I_1(x,t_0+\varepsilon;\{W,\phi(x_0)\})+I_2(x,t_0+\varepsilon;\{W,\phi(x_0)\})  \nonumber\\
&&=I(x,t_0+\varepsilon;\{W,\phi\})
\label{totalequation}
\end{eqnarray}
of which,
\begin{eqnarray}\label{I1equation}
&&I_1(x,t_0+\varepsilon;\{W,\phi(x_0)\}) \nonumber\\
&&=R(x,t_0)\int_{-c}^{+c}{\varepsilon W(x,v,\epsilon)\exp\left(-\mathrm{i}\sqrt{1-v^2/c^2}\varepsilon/\tau_0\right)} \nonumber \\
&&\phi(x+v\varepsilon,t_0)dv ;
\end{eqnarray}
and
\begin{eqnarray}\label{I2equation}
&&I_2(x,t_0+\varepsilon;\{W,\phi(x_0)\})\nonumber\\
&&=R(x,t_0)\int_{c}^{\infty}{\varepsilon W(x,v,\varepsilon)\exp\left(-\sqrt{v^2/c^2-1}\varepsilon/\tau_0\right)} \nonumber\\
&&(\phi(x+v\varepsilon,t_0)+\phi(x-v\varepsilon,t_0))dv.
\end{eqnarray}
Because any smooth function can be expanded into Taylor series, therefore to calculate the function $I(x,t_0+\varepsilon;\{W,\phi\})$, we should calculate the terms $\phi=(x-x_0)^{n}$ firstly.
\begin{eqnarray}\label{ntermI1Eq}
&&I_1(x,t_0+\varepsilon;\{W,(x-x_0)^{2n}\}) \nonumber\\
&&=2R(x,t_0)\tau_0^{2n+1/2}c^{2n+1}\int_{0}^{1}{\varepsilon_0^{2n+1/2}(1-(1-u)^2)^n u^{-1/2}\exp(-\mathrm{i}\varepsilon_0+\mathrm{i}u\varepsilon_0)du} ;
\end{eqnarray}
and,
\begin{eqnarray}\label{ntermI2Eq}
&&I_2(x,t_0+\varepsilon;\{W,(x-x_0)^{2n}\}) \nonumber\\
&&=2R(x,t_0)\tau_0^{2n+1/2}c^{2n+1}\int_{1}^{1+\mathrm{i}\infty}{\varepsilon_0^{2n+1/2}(1-(1-u)^2)^n u^{-1/2}\exp(-\mathrm{i}\varepsilon_0+\mathrm{i}u\varepsilon_0)du},
\end{eqnarray}
where $\varepsilon_0=\varepsilon/\tau_0$.

Considering the contour integral as following
\begin{equation}
\oint f(z)dz=\lim_{a\rightarrow +\infty
}\left(-\int_{1}^{1+\mathrm{i}a}f(y)dy+\int_{1}^{a}f(x)dx+\int_{C\rightarrow
|a|\exp {\mathrm{i}\theta }}f(z)dz\right),
\label{conterI}
\end{equation}
where $f(z)={(1-(1-z)^{2})^{n}z^{-1/2}
\exp (-\mathrm{i}\varepsilon _{0}+\mathrm{i}z\varepsilon _{0})}$. We can see there is
no singular point in the contour area, therefore $\oint f(z)dz=0$. Additionally,
\begin{equation}
\lim_{|z|\rightarrow \infty ,0<{\theta <\pi /2}}f(z)=0,
\end{equation}
consequently
\begin{equation}
\lim_{a\rightarrow \infty }\left(\int_{C\rightarrow |a|\exp {\mathrm{i}\theta }
}f(z)dz\right)=0.
\label{conterI2}
\end{equation}
According to the Eq.~(\ref{conterI}) and Eq.~(\ref{conterI2}), one can find that $\int_{1}^{1+\mathrm{i}\infty}f(y)dy=\int_{1}^{\infty}f(x)dx$.

Therefore, we can evaluate the formula $I$ into
\begin{eqnarray}\label{ntermIEq}
&&I(x,t_0+\varepsilon;\{W,(x-x_0)^{2n}\})  \nonumber\\
&&=2R(x,t_0)\tau_0^{2n+1/2}c^{2n+1}\int_{0}^{+\infty}{\varepsilon_0^{2n+1/2}(1-(1-u)^2)^n u^{-1/2}\exp(-\mathrm{i}\varepsilon_0+\mathrm{i}u\varepsilon_0)du} \nonumber\\
&&=2R(x,t_0)\mathrm{i}^{1/2}\tau_0^{2n+1/2}\exp(-\mathrm{i}\varepsilon_0)c^{2n+1}\Gamma(2n+1/2)\texttt{M}(-n,1/2-2n,2i\varepsilon_0),
\end{eqnarray}
where, $\texttt{M}(a,b,z)$ is Kummer's function of the first kind. Letting $\varphi=\exp(\mathrm{i}px/\hbar)$ in Eq.~(\ref{totalequation}), we show the following relations according discussion above:
\begin{eqnarray}\label{wpEq}
&&I(x,t_0+\varepsilon;\{W,\varphi\}) \nonumber \\
&&=I(x,t_0+\varepsilon;\{W,\exp(-\mathrm{i}px+\mathrm{i}px_0/\hbar)\})\exp(\mathrm{i}px)  \nonumber\\
&&=\sum_{n=0}^{\infty}{I\left(x,t_0+\varepsilon;\left\{W,\frac{(\mathrm{i}p(x-x_0)/\hbar)^{2n}}{2n!}\right\}\right)}\exp(\mathrm{i}px)  \nonumber \\
&&=R(x,t_0)(\mathrm{i}\tau_0\pi)^{1/2}c \left(\frac{1}{\sqrt{1-\mathrm{i}\tau_0pc/\hbar}}+\frac{1}{\sqrt{1+\mathrm{i}\tau_0pc/\hbar}}\right)  \nonumber \\
&&\ \ \ \exp\left(-\mathrm{i}\sqrt{m^2c^4+p^2c^2}\varepsilon/\hbar)\exp(\mathrm{i}px\right).
\end{eqnarray}
To obtain the value of $R(x,t_0)$, we should note that $I(x,t_0;\{W,\varphi\})=\varphi$, therefore, $R(x,t_0)$ should be
\begin{equation}
R(x,t_0)=\hat{R}=\frac{1}{\sqrt{2\mathrm{i}\pi\hbar c^2}}\frac{\hat{H}}{\sqrt{mc^2+\hat{H}}}.
\end{equation}
Then, recalling the combination rule, the infinitesimality $\varepsilon$ in (\ref{wpEq}) can be substituted by arbitrary $t-t_0 $, and the time-evolution equation in one-dimension can be expressed as
\begin{eqnarray}
 \phi(x,t)&=&I(x,t_0+\varepsilon;\{W,\phi\}) \nonumber\\
&=&\int_{-\infty}^{\infty} I(x,t_0+\varepsilon;\{W,\gamma(p)\varphi\})d p \nonumber\\
&=& e^{-\mathrm{i}\frac{t-t_0}{\hbar}\sqrt{m^2c^4+\hat{p}^2c^2}}\int_{-\infty}^{\infty} \varphi\gamma(p)d p \nonumber \\
&=&  e^{-\mathrm{i}\frac{t-t_0}{\hbar}\sqrt{m^2c^4+\hat{p}^2c^2}}\phi(x,t_0).
\end{eqnarray}

It is worth pointing out that Eq.~(\ref{3DTEEq}) is not a Lorentz covariance. This is not surprised because only the positive-energy solution is discussed here and Lorentz invariance could be satisfied if the negative-energy solution is included. Let $\hat{H}^{'}=\sqrt{m^2c^4+(\hat{\vec{p}}-\vec{A}_0)^2c^2}$, and use $\Phi_{+}$ and $\Phi_{-}$ denoting the positive-energy solution and negative-energy solution respectively
\begin{eqnarray}
(\mathrm{i}\hbar\frac{\partial}{\partial t}-V)\Phi_{+}&=&\hat{H}^{'}\Phi_{+} \nonumber\\
(\mathrm{i}\hbar\frac{\partial}{\partial t}-V)\Phi_{-}&=&-\hat{H}^{'}\Phi_{-}.
\label{eqs}
\end{eqnarray}
According to the Eq.~(\ref{eqs}) and letting $\psi_{+}=1/\sqrt{2}(\Phi_{+}+\Phi_{-})$, $\psi_{-}=1/\sqrt{2}(\Phi_{+}-\Phi_{-})$, we get the Lorentz covariant equations as followings
\begin{eqnarray}
(\mathrm{i}\hbar\frac{\partial}{\partial t}-V)^2\psi_{+}&=&(m^2c^4+(\hat{\vec{p}}-\vec{A}_0)^2c^2)\psi_{+} \nonumber\\
(\mathrm{i}\hbar\frac{\partial}{\partial t}-V)^2\psi_{-}&=&(m^2c^4+(\hat{\vec{p}}-\vec{A}_0)^2c^2)\psi_{-}.
\label{covEq}
\end{eqnarray}
A worthy reminder is that although $\psi_{+}$ and $\psi_{-}$ are the two solution of the Klein-Gordon equation, they cannot be used to describe the single particle state. This is because both $\psi_{+}$ and $\psi_{-}$ contain the both positive-energy and negative-energy compositions.

\newcommand{\minitab}[2][l]{\begin{tabular}{#1}#2\end{tabular}}
\begin{table}
\begin{tabular}{|c |c |c |}
  \hline
 Comparison &  EFPI $F(\vec{r},t;\vec{r}_0,t_0)$ & FPI $G(\vec{r},t;\vec{r}_0,t_0)$ \\
  \hline
  Formula Expression & $R(\vec{r},t_0)\sum_{S} W(\wp)\exp(\mathrm{i}S/\hbar)$ &  $C \sum_{S} \exp(\mathrm{i}S/\hbar)$ \\
   \hline
   Low Energy&$\approx G(\vec{r},t;\vec{r}_0,t_0)$  & $G(\vec{r},t;\vec{r}_0,t_0)$ \\
    \hline
  High Energy & $F(\vec{r},t;\vec{r}_0,t_0)$  &  Quantum Field Theory \\
   \hline
  \multirow{2}*{\minitab[c]{Time Evolution}}
& $\mathrm{i}\hbar\partial\phi(\vec{r},t)/(\partial t)=\hat{H} \phi(\vec{r},t)$  & $ \mathrm{i}\hbar\partial\phi(\vec{r},t)/(\partial t)=\hat{H} \phi(\vec{r},t)$ \\
& $\hat{H}=\sqrt{m^2c^4+(\hat{\vec{p}}-\vec{A})^2c^2}+V(\vec{r})$& $ \hat{H}=(\hat{\vec{p}}-\vec{A})^2/(2m)+V(\vec{r})$ \\
 \hline
  \multirow{3}*{\minitab[c]{Relativistic Effect \\and\\ Quantum Nonlocality}}
&\multirow{2}*{\minitab[c]{The relativistic effect of ``path''\\ is presented.}}&\multirow{2}*{\minitab[c]{The relativistic effect of ``path''\\ is unconsidered.}} \\
& &\\
 & $F(x,t_0;x_0,t_0)\neq \delta(x-x_0)$  &  $G(x,t_0;x_0,t_0)= \delta(x-x_0)$ \\
 \hline
 \multirow{3}*{\minitab[c]{Lorentz Covariance?}}
 &\multirow{3}*{\minitab[c]{No. Negative-energy solution is\\ introduced  to keep the covariance. \\ $\longrightarrow$Klein-Gordon Equation.}} &\multirow{3}*{\minitab[c]{N/A. Should be extended \\into quantum field theory}}  \\
 & & \\
 & & \\
 \hline

\end{tabular}
  \caption{The comparison of EFPI to FPI is shown in this table. We can see the quantum nonlocality is discovered when we considerate the relativistic effect of ``paths''. It is namely the quantum nonlocality is due to the relativistic effect in wave function, which have overlooked by us for long time. We also can be seen that the antiparticle appears to be the result of Lorentz covariance according to the EFPI.}
\end{table}

\subsection{Density-flux equation}
It is different from the Klein-Gordon equation which has the second
order of time derivation, wherefore the probability density in Klein-Gordon
theory isn't positive definite and fails to describe a single particle, that Eq.~(\ref{3DTEEq}) is the first order of time derivation and
consequently the probability density defined as $\phi^{\ast}\phi$, which is
the same as the one in the Schr\"{o}dinger theory, is positive definite in EFPI.

In QM, the continuity equation of the probability density describes the transport of conservation of probability density. It is the local form of the conservation law and means that the probability does not increase or decrease when particles move from one place to another. However, because of relativistic effects, the probability density is not locally conserved. It is showed the probability density-flux should obey the following equation
\begin{equation}
\frac{\partial\rho(\vec{r},t)}{\partial t}+\nabla \cdot \vec{j}(\vec{r},t)
+\sum_{n=2}^{\infty}B_n\nabla^nQ_n(\vec{r},t)=0,
\end{equation}
where $\vec{j}(\vec{r},t)=\hbar/(2\mathrm{i}m)Q_1$ and
\begin{eqnarray}
&&Q_n(\vec{r},t)=\phi^{*}(\vec{r},t)\nabla^n \phi(\vec{r},t)-\phi(\vec{r},t)\nabla^n\phi^{*}(\vec{r},t); \nonumber\\ &&B_n=-(-\mathrm{i}\hbar)^{2n-1}c^2n\mathrm{C}_n^{1/2}/(mc^2)^{2n-1},
\end{eqnarray}
of which, $\mathrm{C}_{m}^{n}$ is the binomial coefficient. Here, we call the above equation as density-fluid equation which is different from the original continuity equation since it has the first two terms. It is the last summarizing term, which caused by relativistic effect, that lead to probability density-flux is not a local conserved quantity.

\section{New insight on the wave function collapse and Schr\"{o}dinger cat paradox according to EFPI}

\subsection{The inner correlation of the wave function}
To describe the process of the wave function collapse, we return to the expression of $F$ and for simplification, we only consider the one-dimension situation about $F$. For this situation, we define the operator $\hat{R}$ as
\begin{equation}
\hat{R}=\frac{1}{\sqrt{2\mathrm{i}\pi\hbar c^2}}\frac{\hat{H}^{'}}{\sqrt{mc^2+\hat{H}^{'}}}.
\label{collapseequation}
\end{equation}
We can see $\hat{R}$ and $\hat{H}^{'}$ satisfy the commutation relation, $[\hat{R},\hat{H}^{'}]=0$, therefore, according to the quantum mechanics, the same set of eigenstates $\{\varphi_n\}$ they will be shared. In further, $\{R_n\}$ and $\{E_n\}$ are used to denote the eigenvalues of the operators $\hat{R}$ and $\hat{H}^{'}$ respectively corresponding with the set $\{\phi_n\}$. It is worth detailing the role of the set eigenstates $\{\phi_n\}$ here. In QM, the questions of why the outcome of quantum collapse should be a state of a specific basis and how the specific basis should be chosen for a special measurement puzzle the physiatrists many years. In this text, it is shown that the $\{\phi_n\}$ is the preferred-basis.

We make an assumption here:
\begin{equation}
\hat{H}=\sqrt{m^2c^4+(\hat{p}_x-(A_0(x,t)+A_I(x,t)))^2c^2},
\end{equation}
where $A_0(x,t)$ is the main vector potential and $A_I(x,t)$ is perturbing vector potential, satisfying $|A_I|\ll|A_0|$, and $\langle\langle A_I\rangle\rangle=0$ \cite{note}. $A_I(x,t)$ acts as noise under the main vector potential $A_0(x,t)$. If we regard the main potential $A_0(x,t)$ produced by the apparatus or environment, then $A_I(x,t)$ can be seemed the ``potential noise'', which is produced by the position or movement fluctuation of particles in apparatus or environment. In this article, we consider a simply noise type,
\begin{equation}
A_{I}(x,t)=\sum_{n=0}^{\infty }{f_{n}(\theta (t-n\delta )-\theta
(t-(n-1)\delta ))},
\label{potentialnoise}
\end{equation}
where, the function $\theta (t)$ is the Heaviside step function and $f_{n}$ is the constant function.

For the continuous spectra in one-dimension,
\begin{equation}
F(x,x-\eta;t^{+},t)=\hat{R}\sqrt{\frac{c}{\mathrm{i}|\eta|}}
\exp\left(-\frac{mc|\eta|+\mathrm{i}\int_{x-\eta}^{x}{A(x_0,t)d x_0}}{\hbar}\right).
\end{equation}
It is different from Feynman's theory, in which $K(x,x-\eta;t,t)=\delta(\eta)$, that $F(x,x-\eta;t,t)\neq \delta(\eta)$. Any abrupt local perturbation at $x_0$ of vector potential $A(x_0,t_0) $ \cite{ftnt3} will instantaneously and directly transfers to $x$ though the function $F(x,x_0;t,t)$. This violates the principle of locality that any objective is only influenced directly by its immediate surrounding and the correlation $F(x,x_0;t,t)$ acts as the nonlocality. Hence, the correlation $F(x,x_0;t,t)$ in wave function is nonlocal. If we don't consider the relativistic effect of the paths, this nonlocality will not be discovered. It is indeed the relativistic effect of paths that makes the wave function be bound as a block and much stiffer than previously.

\subsection{Discrete form of wave function collapse equation}
\quad The quantum theory says that any possible state $\phi$ can be expanded into a definite linear combination $\sum_n{a_n\varphi_n}$. Therefore, a unique set of amplitudes $\{a_n\}$ describes a unique state at the basis-space $\{\varphi_n\}$. It is usually supposed that there is no correlation among $\{a_n\}$ except the unitarity equation, but there is a difference about that because the nonlocal character of $F$. It is shown here that the probability amplitudes $a_n$ satisfy the following equations called as the discrete form of wave function collapse equations.
\begin{eqnarray}
a_n(t^+)&=&\sum_{m}{\lambda_{nm}(t^{-}) a_m(t)}; \nonumber \\
\lambda_{nm}(t^-)&=&\langle\varphi_n|\mathfrak{R}(x,t^-)R^{-1}|\varphi_m\rangle; \nonumber \\
\mathfrak{R}(x,t^{-})&=&\frac{\phi(x,t^{-})}{R^{-1}(x,t^{-})\phi(x,t^{-})}.
\label{coefficient}
\end{eqnarray}
Here, $t^+ $ and $t^-$ are used here to distinguish the difference of successive order rather than the time, whereas, in fact, $t^+ =t$ in the above equation \cite{ftnt2}. It is different from the non-relativistical quantum theory since where $\lambda_{nm}$ always equal $\delta_{n,m}
$ that $\lambda_{nm}\neq \delta_{n,m}$ in Eq.~(\ref{coefficient}). The condition which makes $\lambda_{nm}(t)=\delta_{n,m} $ is that $\phi $ is one or a superposition of degenerate states of the eigenstates $\{ \varphi_n\}$. The process of wave function collapse is the one from ``nonlocal'' ($\lambda_{nm}\neq \delta_{n,m}$) to ``local'' ( $\lambda_{nm}= \delta_{n,m}$) under the potential noise $A_I$. Although Eq.~(\ref{coefficient}) is obtained in one-dimension, it is can be extended into three-dimension.

In this case showed in Eq~(\ref{potentialnoise}), ${a_{m}(t)=a_{m}(t}^{-}{
)(1-N_{n}(t^{-}))}$, where
\begin{equation}
N_{n}(t^{-})=\frac{A_{I}(t^{-})c^{2}p_{n}(2mc^{2}+E_{n})}{2E_{n}{}^{2}(mc^{2}+E_{n})%
}.  \label{NoisEq}
\end{equation}

$p_{m}$ is the eigenvalue of $\varphi _{m}$ under the operator $\hat{p}%
-A_{0} $. To deduce the Eq. (\ref{coefficient}) and Eq. (\ref{NoisEq}), we notice that
\begin{eqnarray}
a_{n}(t^{+}) &=&\int_{-\infty }^{\infty }{\varphi _{n}(u,t)^*\phi (u,t)du}
\nonumber \\
&=&\int_{-\infty }^{\infty }{\varphi _{n}(u,t)^* \mathfrak{R}(u,t)
R^{^{\prime }-1}\left(\sum_{m}a_{m}(t^{-})\varphi _{m}(u,t)\right)}.
\label{ampeq}
\end{eqnarray}

If $A_{I}$ is very small, the approximation relation is deduced
\begin{equation}
\hat{R}^{^{\prime }}\approx \hat{R} \left(1-\frac{A_{I}c^{2}(\hat{p}
-A_{0})(2mc^{2}+H^{^{\prime }})}{2(H^{^{\prime }})^{2}(mc^{2}+H^{^{\prime }})
}\right).  \label{appxioR}
\end{equation}
Substitute Eq.~(\ref{appxioR}) into Eq,~(\ref{ampeq}), then the
Eq. (\ref{coefficient}) and Eq. (\ref{NoisEq}) are gotten.

\subsection{The Schr\"{o}dinger cat collapse}
Schr\"{o}dinger cat was a paradox proposed by Sch\"{o}rdinger in his essay to illustrate the ``putative incompleteness'' of quantum mechanics (QM). It is illustrated that a cat live along with a flask containing a poison and a radioactively source. The radioactive source would shatter the flask, releasing the poison and kill the cat, if it decays \cite{Sdg}. Schr\"{o}dinger cat reveals the confliction between the superposition-state description for the behavior of matter on the microscopic level and the definite-status appearance that can be observed on the macroscopic level. Within the standard quantum mechanical formalism, this cat is prepared in a ``mixed state'' -- both dead and alive -- if we don't open the box to look at it. The truth is that this cat never be seen in the macroscopic world. However, the quantum mechanics is experimental proofed to be a very precise theory on the microscopic level up to now. Consequently, determining how to solve the micro-to-macro confliction is still a difficult problem about the physicists.
In this subsection, we will show how to use the EFPI to solve this paradox.

We consider the wave function composition of two eigenstates, which can be written as $\phi (x,t)=a_{0}(t)\varphi
_{0}(x)+a_{1}(t)\varphi _{1}(x)$. Then we get these recursive relations as
following
\begin{equation}
\left(
\begin{array}{c}
a_{0}(t^{+}) \\
a_{1}(t^{+})
\end{array}
\right) =\left(
\begin{array}{cc}
\lambda _{00}(t^{-}) & \frac{a_{0}(t^{-})}{a_{1}(t^{-})}(1-\lambda _{00}(t^{-})) \\
\frac{a_{1}(t^{-})}{a_{0}(t^{-})}(1-\lambda _{11}(t^{-})) & \lambda
_{11}(t^{-})
\end{array}
\right) \left(
\begin{array}{c}
a_{0}(t) \\
a_{1}(t)
\end{array}
\right) .  \label{recursion}
\end{equation}
It is difficult to give an analytical solution about $\lambda _{nm}$ because it
is varying as $\phi $ changing. However, a general estimation of
the value $\lambda _{nm}$ for the two-states wave function can be made. It is noticed that $
\lambda _{00}=R_{1}/R_{0}$ if $
a_{0}^{2}=0$ and $\lambda _{00}=1$ if $a_{0}^{2}=1$. $R_{0}$, $R_{1}$ are the eigenvalues of
$\varphi _{0}$ and $\varphi _{1}$ under the operator $\hat{R}(x,t_{0})$. Obviously, the $\lambda
_{00}$ is the function of $a_{0}^{2}$. The simplest one is linear function,
and then it can be approximately expressed as $\lambda
_{00}=a_{0}^{2}+a_{1}^{2}R_{1}/R_{0}$. Similarly, $\lambda
_{11}=a_{1}^{2}+a_{0}^{2}R_{0}/R_{1}$.

According to the recursive relations, an example is shown in Fig.~1a about the process of collapse of two-states, which can be seen as the cat's state $a_0|\textmd{alive}\rangle+a_1|\textmd{dead}\rangle$ \cite{ftnt0}, varying under the action of the potential noise $A_I$. One can see at initial stages these lines of the two amplitudes are seriously oscillatory, but they calm down as soon as one of them become zero (one) and then the collapses finish. One type of potential noise is shown in Fig.~1b. The time of these processes, which is proportional to $\delta$ (the character of the potential noise $A_I$) are very short. Actually, the value of $\delta$ we proposed in Fig.~1b is a rough estimate one here \cite{ftnt1}, and it is believed much shorter in real world, therefore the process of collapse is instantaneous. According to Fig.~1, it can be seen that the different sets of potential noise $A_I$ cause the different results of wave function collapse. As many numerical calculations are made, we find that under the different sets of noise, the ratio $|\textmd{alive}\rangle$ to $|\textmd{dead}\rangle$ approach $a_0^2:a_1^2$, which is the prediction (Burn-rule) according to quantum theory.
\begin{figure}
  \includegraphics[width=0.7\textwidth]{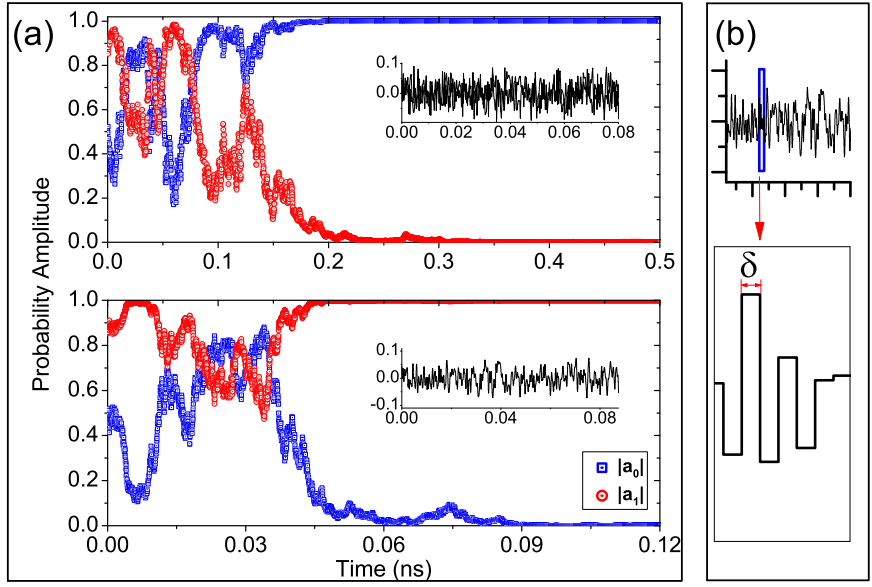}\\
  \caption{The process of collapse and the ``potential-noise''. (a) shows the process of collapse under the potential-noise. The red line denotes the absolute value of probability amplitude $a_0(t) $ with the initial value $1/2 $, and blue line denotes $a_1(t) $ with the initial value $\sqrt{3}/2 $. The black oscillatory lines in windows are the sets of noise function $A_I(x,t) $. The different set of noise causes the different result of collapse; and the time of collapse are $0.3$ ns in the top picture and $0.1$ ns in the bottom picture. (b), the top oscillatory curve is the function $A_I(x,t)$ with time and the bottom one is the zoom of the blue rectangular region of top one. For this type, the function of $A_I(x,t)$ can be expressed as $\sum_n{f_n\left(\theta(t-n\delta)-\theta(t-(n+1)\delta\right)}$, in which $\theta(t)$ is the Heaviside step function whose value is zero for negative argument and one for positive argument. The $\delta$ is chosen as $0.25*10^{-13}{\rm ns}$ and $E_0/E_1=1.25/1.75$.}\label{Fig1}
\end{figure}

Therefore, the basis-space where collapse will happen is determined by the main potential $A_0$ \cite{ftnt10,zeff} and the tiny potential noise $A_I$ chooses the outcome of quantum measurement in this basis-space. It is because the nonlocal correlation $F(x,x_0;t,t)$ is hidden in the wave function that the noise $A_I$ is no longer an useless and unwanted role in physics. The fluently and uncontrolled potential noise $A_I$ makes our world more definite and surprising. Why does the collapse relate the nonlocality? It can be understood as this: If there is no nonlocal correlation in wave function, the influence of ``potential-noise'' will be counteracted by its serious oscillatory because the action of every point of noise can't transport instantaneously; Oppositely, if the nonlocal correlation exists, the actions of ``potential-noise'' will be accumulated because the nonlocal correlation can redistribute these actions to whole wave function instantaneously, and then the wave function will finally collapse into a stable state, namely $\varphi_n $ in the article. This is why the prefer-basis exists.

Moreover, we can give a difference between ``measurement'' between ``operation''. The ``measurement'' happen on condition that the interaction of system-environment should be big enough to distinguish the eigenstates $\{ \varphi_n\}$. If the interaction is not big enough and $R_n\approx R_m $, then $\lambda_{mn}=0$ according to Eq.~(\ref{coefficient}) and collapse will not happen. In addition, the ``measurement'' also needs ``potential-noise'' to participate in, so the instruments of measurement should be ``macro'' enough to produce enough noise. Conversely, the realization of ``operation'' should suppress potential-noise. Therefore, the Schr\"{o}dinger cat will be immediately collapse to be dead or alive in the noise world granted it really exists. Hence, we always ``see'' the world in definite and determinate.

\section{Prediction and Experiment suggestion}
Finally, we briefly mention a possible experiment to test and valify this theory. In the experiment to detect the magnetic Aharonov-Bohm effect,  the phase $\varphi_{L_1}$ of electrons that go through the left side of the solenoid has different value with the one $\varphi_{L_2}$ that go through the right side. $\varphi_{L_1}-\varphi_{L_2}=-e\int_S \vec{B_0}\cdot d\vec{S}/\hbar$. It is known as the Aharonov-Bohm phase. However, if we add a fluctuated magnetic field $\vec{B}_1$, which is much less than $\vec{B_0}$ and satisfied $\int_{t}^{t+\tau}\vec{B_1}dt^{'}=0$ ($\tau$ is the time that the electrons go through the distant between slits to the electron-detector), we could deduce the different phenomenons between the previously theory and EFPI. According to the previously theory in QM, this fluctuated magnetic field $\vec{B}_1$ should have no effect for the Aharonov-Bohm phase and the interference fringe will not destroy. However, in EFPI, the interference fringe would be expected to be destroyed. This is because the nonlocal correlation of wave function.

Because the movement of the electron on the left side has the same direction with the vector potential and the right one has the contrary direction, therefore the main Hamiltonian on the left is $\sqrt{m^2c^4+(\hat{p}+eA_0)^2c^2}$ and the right one is $\sqrt{m^2c^4+(\hat{p}-eA_0)^2c^2}$. Consequently, it forms a two-state system and can be analysis with the above conclusion. If the fluctuated magnetic field added into as the form
\begin{equation}
\vec{B}_1=\sum_{n=0}^{\infty }{(-1)^n\vec{B}_{cons}(\theta (t-n\delta )-\theta
(t-(n-1)\delta ))},
\end{equation}
where, $\delta\ll\tau$. Under this fluctuated field, electrons will tend to go through the left slit. Then the interference effect will be disappearing and diffraction effect will come forth.

To realize this experiment, we suggest it is worked in the graphene because the effective mass of electron is zero when electrons are in the Dirac point. Hence, we can get the distinct difference between this two slits without adding a big magnetic field into.

\section{Conclusions }
Measurement, in quantum theory, is not just only a theory concerning the Schr\"{o}dinger cat in alive or dead, or the moon being here or not, but also the key and basement to the problem of the interpretation of QM. In fact, the different views for quantum measurement yield different interpretation for QM, such as the Copenhagen interpretation, relative-state interpretation, Bohmian mechanics and so on. It has attracted many attentions of physicists since the beginning of quantum theory establishment, but there is still no consensus. The measurement problem blocks up the way to dig out the value of QM especially in recent years when the applications of quantum theory are gradually expanded. It is known that the local quantum measurement under one of the two distanced entangled particle will cause the corresponding state changing for the other one but the quantum unitary transformation will not. The important thing is that the question of how to distinguish quantum measurement and quantum unitary transformation in a real experiment is unknown yet. In fact, The spook action used by Albert Einstein to denote the unique character of the entangled particle is mainly due to the nonlocal effect of quantum measurement.

In this paper, we discuss the extension of the Feynman path integral. According to this extension we get not only the new time-evolution equation but also the wave function collapse equation. Different from quantum field theory, we just analyze the inner correlation of wave function rather than the interaction among fields in the relativity realm and therefore we show an inner mechanism of wave function that has never been discovered before. It is also different from the decoherence theory although they both consider the influence of environment. Our theory pays more attention to potential influence of environment whereas the decoherence theory pays more attention to state-decoherence under the influence of environment and it is still a direct application of the non-relativistical quantum theory \cite{us2,us3}. Additionally, comparing the Dynamical Reduction Models \cite{us5,RDM1,RDM2,RDM3,RDM4}, which is the nonlinear revised for Schr\"{o}dinger equation, our theory is a distinct different approach and show this theory can solve not only the preferred-basis problem but the question of why does the collapse have the instantaneous and stochastic properties. We believe this theory is a way, or granted it is not, it support a new orientation, to solve the measurement problem.

\section{Acknowledgements}
This work was supported by the National Basic Research Program of China (973 Program) grant No. G2009CB929300 and the National Natural Science Foundation of China under Grant Nos. 10874013, 60776061 and 60821061.


\begin{thebibliography}{00}
\bibitem{ftnt3} Because the non-relativistic action is the quadratic term, it can be easily deal with the integral computation. Moreover, Feynman deduced the Schr\"odinger evolution equation basing on the non-relativistic action.
\bibitem{ftnt3} it seems that it doesn't relate with the scalar potential energy $V(x,t)$. In fact, the impact of scalar potential is contained into vector potential. From the gauge transformation, we know, $A^{'}(x,t)=A(x,t)+\int^{t}{\frac{\partial U(x,t)}{\partial x}dt} $. Therefore, $A^{'}(x,t^+)-A^{'}(x,t)=A(x,t^+)-A(x,t)+\int_t^{t^+}\frac{\partial U(x,t)}{\partial x}dt $. The last term of the equation equal zero if $U(x,t) $ is not a singular function.
\bibitem{note} $\langle\langle \cdot\rangle\rangle$ is used to denote the average value of the quantity contained within the ``brackets''.
\bibitem{ftnt2} In math, we usually use the $f^{'}(x+0) $ denote the derivation on the right side of point $x $. So we use the similar notation in this article. $t^+=t+0$ denotes the right side of $t$, and $t^-$ denotes the left side of it. But the values of them are the same.
\bibitem{ftnt0}  In fact, to strictly speak, the cat's state is an entanglement state with radioactive substance. It is usually expressed as $a_0|\mathrm{alive}\rangle\otimes|\wedge\rangle+a_1|\mathrm{dead}\rangle\otimes|\vee\rangle$, where $\vee$ and $\wedge$ denotes the Geiger counter detects the radiation and doesn't detects the radiation respectively. In this article, we leave out the subspace of the radioactive substance because this subspace is not a focal point for us to discussion the wave function collapse.
\bibitem{ftnt1}  In fact, we can give an estimative value about the time of this collapse. The frequency "potential-noise" of measurement -structure can be approximately written as $\omega=k_BT/\hbar$. So at room temperature, $\delta\approx1/\omega\approx0.25*10^{-13}$ \textmd{s}, which is also the value we using in this article.
\bibitem{ftnt10} We notify that the states $\{\varphi_n\} $ are not the eigenstates of the total Hamiltonian $\hat{H}=\hat{H}{'}+V(x,t)$ if $V(x,t)\neq 0 $, but the effect of the system evolution can be overlooked. This is because the production of potential-noise is continuous, and in fact causes the continuous observation processes on system, which has been talked in theory of Zeno effect.
\end{thebibliography}
\end{document}